\documentclass[aps,prl,twocolumn,superscriptaddress,groupedaddress]{revtex4}  

\usepackage{graphicx}  
\usepackage{url}                  
\usepackage{dcolumn}       
\usepackage{bm}                  
\usepackage{amssymb}       
\usepackage{mathtools}     
\DeclarePairedDelimiter\ket{\lvert}{\rangle}
\usepackage{amsmath}    
\usepackage{pgf}                   
\usepackage{tikz,pgfplots}                   
\usetikzlibrary{shapes,arrows,calc}
\usetikzlibrary{arrows, decorations.markings}
\usetikzlibrary{decorations.pathreplacing,calligraphy}

\begin{document}  

\title{Multidimensional Quantum Fourier Transformation}
\author{Philipp Pfeffer}
\affiliation{Institute of Thermodynamics and Fluid Mechanics, Technische Universit\"at Ilmenau, Postfach 100565, D-98684 Ilmenau, Germany}
\date{\today}

\begin{abstract}
 The Quantum Fourier Transformation (QFT) is a well-known subroutine for algorithms on qubit-based universal quantum computers. In this work,  the known QFT circuit is used to derive an efficient circuit for the  multidimensional QFT. The complexity of the algorithm is $\mathcal{O}(  \log^2(M)/d )$ for an array with $M=(2^n)^d$ elements $(n \in \mathbb{N})$ equally separated along $d$ dimensions. Relevant properties for application are discussed. An example on current hardware is depicted by a 6 qubit 2D-QFT with an IBM quantum computer.
\end{abstract}
\maketitle

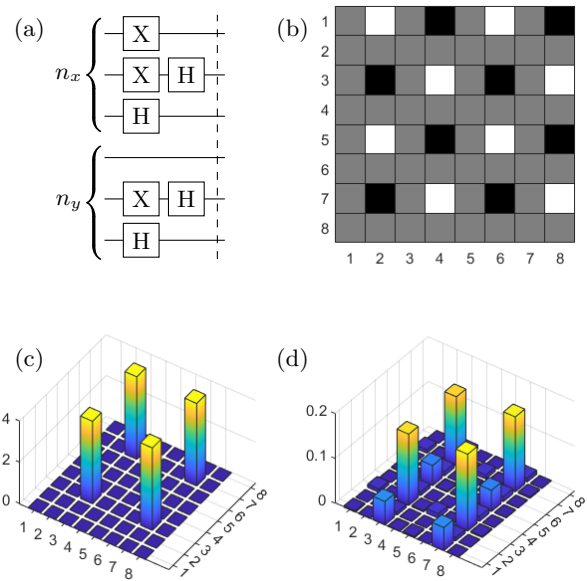
\begin{figure}[h]
    \centering
    \begin{tikzpicture}

        \draw[] (-1.5,-4.7) node[anchor=south west]{\includegraphics[scale=0.5]{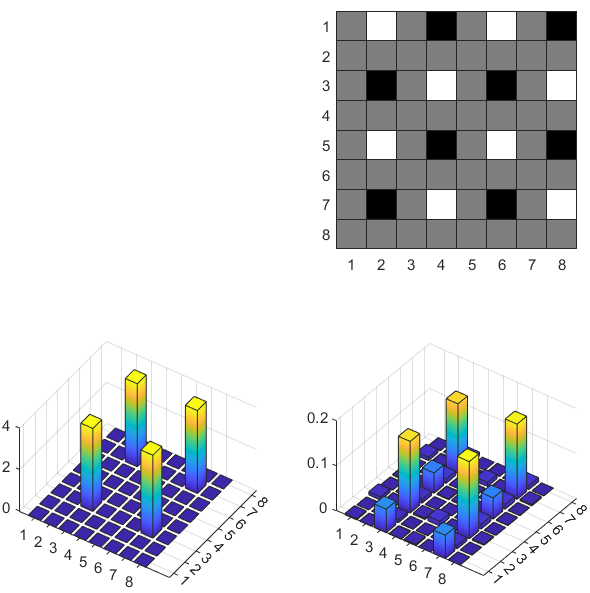}};
        \def\y{0.55};
        \draw[] (0,0) -- (1.6,0);
        \foreach \r in {1,...,5}{
            \draw (0,\r*\y) -- (1.6,\r*\y);
        }
        \draw[fill=white] (0.25,-0.225) node[anchor=south west]{H} rectangle ++(0.475,0.45);
        \draw[fill=white] (0.25,0.325) node[anchor=south west]{X} rectangle ++(0.475,0.45);
        \draw[fill=white] (0.85,0.325) node[anchor=south west]{H} rectangle ++(0.475,0.45);
        \draw[fill=white] (0.25,1.425) node[anchor=south west]{H} rectangle ++(0.475,0.45);
        \draw[fill=white] (0.25,1.975) node[anchor=south west]{X} rectangle ++(0.475,0.45);
        \draw[fill=white] (0.85,1.975) node[anchor=south west]{H} rectangle ++(0.475,0.45);
        \draw[fill=white] (0.25,2.525) node[anchor=south west]{X} rectangle ++(0.475,0.45);

        \draw[dashed] (1.5,-0.25) -- ++(0,3.25);

        \draw[decoration={calligraphic brace,amplitude=5pt}, decorate, line width=1.25pt] (-0.055,1.45) -- ++(0,1.5) ++(0,-0.75) node[anchor =  east] {$n_x$ \ } ;
        \draw[decoration={calligraphic brace,amplitude=5pt}, decorate, line width=1.25pt] (-0.055,-0.25) -- ++(0,1.5) ++(0,-0.75) node[anchor =  east] {$n_y$ \ } ;
        \draw[] (-1,2.8) node[]{(a)};
        \draw[] (-1,-1.6) node[]{(c)};
        \draw[] (2.5,2.8) node[]{(b)};
        \draw[] (2.5,-1.6) node[]{(d)};
    \end{tikzpicture} 
    \caption{Example of a 2D-QFT on the IBM quantum computer \textit{'ibm\_lagos'}. (a) represents the quantum circuit that initializes $f(x,y)$ of Eq. \eqref{Eq:fx}. (b) shows the corresponding initialized image. (c) illustrates the absolute values of the classical 2D-FFT. (d) shows the result extracted from \textit{'ibm\_lagos'} by $2^{14}$ samples.}
    \label{fig:Exp}
\end{figure}

The Quantum Fourier Transformation \cite{QFT} (QFT) is a key subroutine in quantum information processing, most prominently used within the quantum phase estimation \cite{Kit_QPE} and the factoring algorithm of Shor \cite{Shor_F}. Comparing to its classical counterpart, the fast fourier transformation \cite{FFT} (FFT) solves the same problem with effort $\mathcal{O}(N\log(N))$ while the QFT needs  $\mathcal{O}(\log^2(N))$ operations \cite{NielsenChua} for a vector with $N=2^n$ elements ($n \in \mathbb{N}$). Though this speed-up does not lead to the broad replacement of the FFT, the aspect of quantum parallelism is important for the construction of further quantum algorithms. Single operations acting on a large quantum state, such that the whole state is affected, enable the speed-up of the QFT and can be utilized even better for a multidimensional QFT. \\
The idea of this dimensional extension is not new, as the 2-dimensional QFT (2D-QFT) is of crucial use in the field of quantum image processing \cite{QIP,QIP_2,QIP_3}. There, it can be used for edge detection \cite{QIP_Edge}, watermarking \cite{QIP_Watermark} and for the implementation of a discrete cosine transform \cite{QCoT}, which is useful for interpolation \cite{QIP_Inter}. The motivation here is to extend the concept to more dimensions and give an easily understandable summary of the constructed circuit and its complexity. Furthermore, this work aims at readers new to the field in order to help them understand the concept of quantum parallelism as well as the consequences hindering its success. \\ 
The exact task is to construct a quantum circuit which calculates the d-dimensional discrete Fourier transformation of a d-dimensional array A where each dimension $i$ spreads over $N_i=2^{n_i}$ elements ($n_i \in \mathbb{N}$). The transformation is given by the formula
\begin{equation}
    \tilde{a}_{\delta_1,..,\delta_d} = \sum\limits_{k_d=0}^{N_d-1}  \omega_{N_d}^{k_d \delta_d} \sum\limits_{k_{d-1}=0}^{N_{d-1}-1}  \cdots \left( \sum\limits_{k_1=0}^{N_1-1} \omega_{N_1}^{k_1\delta_1} a_{k_1,..,k_d}  \right) 
    \label{Eq:MDFT}
\end{equation}
 with $\omega_{N_i}=e^{j2\pi/N_i}$ and $a_{k_1,...,k_d}$ as the indexed elements of A. For simplification, indices are abbreviated by
$$ N_1\cdot...\cdot N_{i-1}=:N_{i_\downarrow } \ \ , \ \  N_{ i^\uparrow}:= N_{i+1}\cdot... \cdot N_{d}$$
and the number of all array elements is $M=N_1\cdot ... \cdot N_d$. \\ \ \\
The 1-dimensional FFT has a matrix representation given by the Vandermonde-matrix
\begin{equation*}
    V_{N_i} = \begin{pmatrix}
            \omega_{N_i}^0 & \omega_{N_i}^0 & \omega_{N_i}^0 & \cdots & \omega_{N_i}^0 \\
            \omega_{N_i}^0 & \omega_{N_i} & \omega_{N_i}^2 & \cdots & \omega_{N_i}^{N_i-1} \\
            \omega_{N_i}^0 & \omega_{N_i}^2 & \omega_{N_i}^4 & \cdots & \omega_{N_i}^{2(N_i-1)} \\
            \vdots & \vdots & \vdots & \ddots & \vdots \\
            \omega_{N_i}^0 & \omega_{N_i}^{N_i-1} & \omega_{N_i}^{2(N_i-1)} & \cdots & \omega_{N_i}^{\left( (N_i-1)^2\right)}
    \end{pmatrix}.
\end{equation*}
The QFT in terms of a matrix is the rescaled version of this matrix, namely
\begin{equation*}
    QFT_{N_i} = \frac{1}{\sqrt{N_i}} V_{N_i}.
\end{equation*}
Thereby the matrix is unitary. The circuit implementing this matrix (see \cite{NielsenChua}) will be called $QFT_{n_i}$. Note that the prefactor of $QFT_{N_i}$ matters for comparing the results with the classical algorithm, as done in Fig. \ref{fig:Exp}.  \\
In classical computing, multidmensional FFT algo\-rithms utilize the FFT recursively \cite{FFTW3}. For the solution, one can start by applying the FFT to the last sum of Eq. \eqref{Eq:MDFT} to get an interim array where every entry is transformed along the first dimension. This array is then the input for the FFT along the next dimension, which can be repeated until every sum is evaluated. The $d=2$ case gives this procedure the name row-column algorithm, since there one transforms the array first along its rows or columns in order to transform along the other next. This idea is essential for the following algorithm.   \\

\begin{figure}[h]
    \centering
    \Large
    \begin{tikzpicture}[scale=1.2]
            
            \def\y{0};
            \draw[] (0,2.5+\y) -- ++(4,0);
            \draw[] (0,2.65+\y) -- ++(4,0);
            \draw[] (-0.025 ,2.35+\y) node[anchor=west]{
            \begin{tikzpicture}[scale=0.3]
                \draw[] (0,0) -- ++(0,0.1);    \draw[] (5,0) -- ++(0,0.1);
                \draw[] (0,0.25) -- ++(0,0.1); \draw[] (5,0.25) -- ++(0,0.1);
                \draw[] (0,0.5) -- ++(0,0.1);  \draw[] (5,0.5) -- ++(0,0.1);
            \end{tikzpicture}};
            \draw[] (1.75 ,2.35+\y) node[anchor=west]{
            \begin{tikzpicture}[scale=0.3]
                \draw[] (0,0) -- ++(0,0.1);    \draw[] (8,0) -- ++(0,0.1);
                \draw[] (0,0.25) -- ++(0,0.1); \draw[] (8,0.25) -- ++(0,0.1);
                \draw[] (0,0.5) -- ++(0,0.1);  \draw[] (8,0.5) -- ++(0,0.1);
            \end{tikzpicture}};
            \draw[] (0,2.2+\y) -- ++(4,0);
            \draw[fill=white] (2.25,2.1+\y) node[anchor=south west]{$QFT_{n_1}$} rectangle ++(1.5,0.65);
            
            \def\y{-0.8};
            \draw[] (0,2.5+\y) -- ++(4,0);
            \draw[] (0,2.65+\y) -- ++(4,0);
            \draw[] (-0.025 ,2.35+\y) node[anchor=west]{
            \begin{tikzpicture}[scale=0.3]
                \draw[] (0,0) -- ++(0,0.1);    \draw[] (5,0) -- ++(0,0.1);
                \draw[] (0,0.25) -- ++(0,0.1); \draw[] (5,0.25) -- ++(0,0.1);
                \draw[] (0,0.5) -- ++(0,0.1);  \draw[] (5,0.5) -- ++(0,0.1);
            \end{tikzpicture}};
            \draw[] (1.75 ,2.35+\y) node[anchor=west]{
            \begin{tikzpicture}[scale=0.3]
                \draw[] (0,0) -- ++(0,0.1);    \draw[] (8,0) -- ++(0,0.1);
                \draw[] (0,0.25) -- ++(0,0.1); \draw[] (8,0.25) -- ++(0,0.1);
                \draw[] (0,0.5) -- ++(0,0.1);  \draw[] (8,0.5) -- ++(0,0.1);
            \end{tikzpicture}};
            \draw[] (0,2.2+\y) -- ++(4,0);
            \draw[fill=white] (2.25,2.1+\y) node[anchor=south west]{$QFT_{n_2}$} rectangle ++(1.5,0.65);

            \def\y{-2.2};
            \draw[] (0,2.5+\y) -- ++(4,0);
            \draw[] (0,2.65+\y) -- ++(4,0);
            \draw[] (-0.025 ,2.35+\y) node[anchor=west]{
            \begin{tikzpicture}[scale=0.3]
                \draw[] (0,0) -- ++(0,0.1);    \draw[] (5,0) -- ++(0,0.1);
                \draw[] (0,0.25) -- ++(0,0.1); \draw[] (5,0.25) -- ++(0,0.1);
                \draw[] (0,0.5) -- ++(0,0.1);  \draw[] (5,0.5) -- ++(0,0.1);
            \end{tikzpicture}};
            \draw[] (1.75 ,2.35+\y) node[anchor=west]{
            \begin{tikzpicture}[scale=0.3]
                \draw[] (0,0) -- ++(0,0.1);    \draw[] (8,0) -- ++(0,0.1);
                \draw[] (0,0.25) -- ++(0,0.1); \draw[] (8,0.25) -- ++(0,0.1);
                \draw[] (0,0.5) -- ++(0,0.1);  \draw[] (8,0.5) -- ++(0,0.1);
            \end{tikzpicture}};
            \draw[] (0,2.2+\y) -- ++(4,0);
            \draw[fill=white] (2.25,2.1+\y) node[anchor=south west]{$QFT_{n_d}$} rectangle ++(1.5,0.65);
            
            \draw[] (0.06 ,1) node[]{$\vdots$};
            \draw[] (1.55 ,1) node[]{$\vdots$};
            \draw[] (3 ,1) node[]{$\vdots$};

            \draw[fill=white] (0.25,-0.25) node[anchor=north west, align=left,rotate=90]{\ \ initialize($\ket{v}$)} rectangle ++(0.75,3);

            \draw[decoration={calligraphic brace,amplitude=5pt}, decorate, line width=1.25pt] (-0.055,2.1) node[anchor = south east] {$n_1$ \ } -- (-0.055,2.7);
            \draw[decoration={calligraphic brace,amplitude=5pt}, decorate, line width=1.25pt] (-0.055,1.25) node[anchor = south east] {$n_2$ \ } -- (-0.055,1.95);
            \draw[decoration={calligraphic brace,amplitude=5pt}, decorate, line width=1.25pt] (-0.055,-0.1) node[anchor = south east] {$n_d$ \ } -- (-0.055,0.5);
            
            \draw[decoration={calligraphic brace,amplitude=5pt}, decorate, line width=1.25pt] (4.2,2.7) -- (4.2,-0.1);
            \draw[] (4.3,1.3) node[anchor =  west] {$\ket{\hat{v}}$};
    \end{tikzpicture}
    \caption{Quantum circuit to calculate the d-dimensional FFT. $\ket{v}$ represents the input array and $\ket{\hat{v}}$ the QFT of this array, both as quantum states.}
    \label{fig:Circ}
\end{figure}
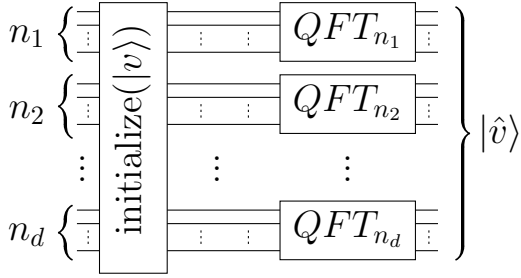

The quantum circuit for the multidimensional QFT is shown in Fig. \ref{fig:Circ} and works as follows. The elements of $A$ can be aligned into a vector 
$$\vec{v} = (a_{1,1,..,1}\ ,\ ...\ , \ a_{N_1,1,..,1}\ ,\ a_{1,2,..,1}\ , \ ... \ , \ a_{N_1,N_2..,N_d} )^T.  $$
This vector has to be reformulated into a quantum state $\ket{v}$, which requires to scale the vector by
\begin{equation}
    \ket{v} = \frac{\vec{v}}{\sqrt{\sum\limits_{k=1}^{N_1 \cdot ... \cdot N_d} |v_k|^2}}.
    \label{Eq:Scale}
\end{equation} 
With that $\ket{v}$ can be initialized on a quantum computer by a state loading procedure\cite{State_Ini,State_Ini2}. As pointed out later on, it is more favourable if the input construction already is an efficient algorithm for a specific problem. \\
 
The formalism for circuit-to-matrix conversion is that the upper qubit is the least significant bit, or in other words the last bit of the bit string. Thereby, if the action of $QFT_{n_1}$ in Fig. \ref{fig:Circ} is of interest, the corresponding matrix acting on the input state is by definition \cite{NielsenChua} 
\begin{equation}
U_1 = \mathcal{I}_{N_{1^\uparrow}} \otimes QFT_{N_1} = \begin{pmatrix} QFT_{N_1} &   & \mathbb{O} \\
  \  & \ddots &  \\
\mathbb{O} &   & QFT_{N_1} \\
\end{pmatrix}  
\label{Eq:U1}
\end{equation}
where $\mathcal{I}_N$ denotes the N$\times$N identity matrix, $\mathbb{O}$ represents all zeros besides the block diagonal and $\otimes$ is the tensor product. The tensor product is the mathematical representation of quantum parallelism since it duplicates the operation to the size of the quantum state it acts on. As a result, this circuit gives
$$ U_1\ket{v} = \begin{pmatrix}
    QFT_{N_1}\    (\ v_1\ ,\ ...\ ,\ v_{N_1}\ )^T    \\
    QFT_{N_1}\ (v_{N_1+1},...,v_{2N_1})^T \\
    \vdots \\
     QFT_{_1}\ (v_{M-N_1},...,v_{M})^T
\end{pmatrix} $$
which creates the first interim matrix comparable to the classical approach. In other words, a single QFT finishes the first dimension globally. In general for any dimension i, the large unitary matrix corresponding to the quantum circuit and its position in Fig. \ref{fig:Circ} is
\begin{equation}
U_i = \mathcal{I}_{N_{i^\uparrow}} \otimes \underbrace{QFT_{N_i} \otimes \mathcal{I}_{N_{i_\downarrow}}}_{\tilde{U}_i}.
\label{Eq:Ui}
\end{equation}
For edge cases, $\mathcal{I}_{1_\downarrow} =\mathcal{I}_{d^\uparrow} = 1$ is the empty product convention. The arrangement of A as a vector in combination with Eq. \eqref{Eq:Ui} yields the transformation along dimension i. To further clarify this, writing out $\Tilde{U}_i$ of Eq. \eqref{Eq:Ui} by the tensor product definition gives

\begin{equation}
    \Tilde{U}_i = \frac{1}{\sqrt{N_i}} \begin{pmatrix}
        \omega^0_{N_i} \ \mathcal{I}_{N_{i_\downarrow}} & \omega^0_{N_i} \mathcal{I}_{N_{i_\downarrow}}  & ... & \omega^0_{N_i} \mathcal{I}_{N_{i_\downarrow}} \\
        \omega^0_{N_i} \mathcal{I}_{N_{i_\downarrow}}  & \omega^1_{N_i}\mathcal{I}_{N_{i_\downarrow}} & ... & \omega^{N_i-1}_{N_i} \mathcal{I}_{N_{i_\downarrow}} \\
        \vdots & \vdots & \ddots & \vdots \\
        \omega^0_{N_i} \mathcal{I}_{N_{i_\downarrow}} & \omega^{N_i-1}_{N_i} \mathcal{I}_{N_{i_\downarrow}} & ... & \omega_{N_i}^{(N_i-1)^2} \mathcal{I}_{N_{i_\downarrow}}
\end{pmatrix}.
\label{Eq:Ui_Ti}
\end{equation}

The overall working principle of the algorithm can be explained by observing Eq. \eqref{Eq:Ui} with the help of Eq. \eqref{Eq:U1} and Eq. \eqref{Eq:Ui_Ti}. The position of $QFT_{n_i}$ in the circuit structure of Fig \ref{fig:Circ} forces the tensor product of Eq. \eqref{Eq:Ui}.  The evaluation of the latter tensor product, namely for $\Tilde{U}_i$, creates a matrix as shown in Eq. \eqref{Eq:Ui_Ti}. Here, the elem\-ents of the 1-dimensional QFT are separated such that, combined with the vector form of the original array, they act only along the dimension i. The first tensor product, similar to Eq. \eqref{Eq:U1}, then duplicates this correctly spaced matrix to the necessary size in a block diagonal structure, which assures the action on all array elements. All QFTs act on separate qubits, which means that all $U_i$ of Eq. \eqref{Eq:Ui} commute with each other \cite{NielsenChua}. Thereby, one can choose that they act in ascending order, i.e. they evaluate the sums as done in Eq. \eqref{Eq:MDFT}, dimension by dimension, similar to the classical algorithm. To summarize, each QFT transforms the array along the corresponding dimension with a single QFT while all QFTs are executed simultaneously.\\

The difference of computational complexity is thereby significant. For simplicity, let A be a d-dimensional array equally sized along every dimension i such that every dimension has $N_i=N=2^n$ elements and $M=N^d$. It is knwon for the classical multidimensional FFT that it has computational effort $\mathcal{O}{(M\log(M))}$ \cite{FFTW3}, similar to the 1-dimensional case. For the quantum algorithm, a single QFT has a known complexity of $\mathcal{O}(n^2)$, and the algorithm needs d QFTs in total, so $\mathcal{O}(d n^2)$ operations. In terms of $M$, one can deduce that $n = \log_2(\sqrt[\leftroot{1} \uproot{2} d ]{M})$ which means for a direct comparison that
$$\text{FFT:} \  \mathcal{O}{(M \log(M))} \leftrightarrow \text{QFT:} \ \mathcal{O}( \log^2(M)/d ).$$
The applicability of this speed up remains to be shown since it is already known for the $d=1$ case that this algorithm will not replace all applications of the FFT \cite{NielsenChua}. The computational effort of rescaling arrays according to Eq. \eqref{Eq:Scale}, then initializing them on a quantum computer \cite{State_Ini,State_Ini2} in order to approximate the spectrum with measurements will almost always scale at least linear in terms of array elements $M$, which makes the computational advantage either shallow or non-existing. Additionally, the approximation of the final spectrum by measurements will return $|\hat{v}_i|^2$, which will erase both the sign as well as the phase information of the complex Fourier coefficient $\hat{v}_i$. These points have to be addressed in competition with the classical multidimensional FFT, which is efficiently executable with parallel computing approaches \cite{FFTW3,Opt_FFT}.  \\
In favor of the proposed algorithm speak the existing applications of QFTs, as is already the case in the field of quantum image processing \cite{QIP_Edge,QIP_Inter,QIP_Watermark,QCoT}. Another application could use the multidimensional QFT to solve Laplace operators by variational quantum algorithms (see \cite{VQA_Lubasch}) with more dimensions. Furthermore, it is worth noting that the algorithm itself does not scale with the amount of parallel initialized arrays. In detail, if one manages to initialize not only a single array $M$ as $\ket{v}$ done by Eq. \eqref{Eq:Scale} but multiple ones aligned one after another, then the circuit of Fig. \ref{fig:Circ} can be appended where only the initialization spreads over more qubits. In the herein notation, these qubits are added below the illustrated circuit, which then creates a structure similar to Eq. \eqref{Eq:U1} whereas the 1-dimensional QFT is replaced by the d-dimensional QFT. \\
Concluding, an optimal use-case for this algorithm would be that multiple input arrays need to be calculated first, and it exists an efficient quantum algorithm to do so. The result state after the QFT should then either be strongly localized or further processible on a quantum computer, for instance if the difference of multiple Fourier transformed arrays is of interest. \\

As a final example, a real quantum device of IBM (\textit{'ibm\_lagos'}) is used to calculate the Fourier coefficients of an 8$\times$8 picture ($d=2$ and $n_x=n_y=3$), where two  simplifications are used. First, the final swap operations in the QFTs (see \cite{NielsenChua}) are avoided by measuring the qubits in a swapped order. Further circuit optimization has been inhibited rigorously. Second, the input picture is generated by 
\begin{equation}
    f(x,y) = sin(\pi x/2) cos(\pi y/2 )
    \label{Eq:fx}
\end{equation}
where $x,y \in \{0,1,...7\}$. This is easy to initialize, as Fig. \ref{fig:Exp}.a shows. The image itself is illustrated in Fig. \ref{fig:Exp}.b. For this, four distinct peaks are expected in the Fourier spectrum, as depicted in Fig. \ref{fig:Exp}.c. The result of the real quantum computer, shown in Fig. \ref{fig:Exp}.d , has its largest peaks at the correct locations. Here, the expectable influence of noisy quantum computers corrupts the result such that the peaks are not equally sized and additional peaks rise. The lowest correct and highest incorrect peak still show  significant difference.

\ \\
\acknowledgments
Closing with acknowledgments, I like to thank Theo Käufer for bringing up the fundamental motivation for this work. I thank Darius Becher and Clara Stolzenberg for helpful comments. During this work, I was supported by the project no. P2018-02-001 "DeepTurb -- Deep Learning in and of Turbulence" of the Carl Zeiss Foundation and the Deutsche Forschungsgemeinschaft under grant no. DFG-SPP 1881. \\
I acknowledge the use of IBM Quantum services for this work. The views expressed are those of the author, and do not reflect the official policy or position of IBM or the IBM Quantum team. In this paper {\em 'ibm\_lagos'} was used, which is an IBM Quantum Falcon Processor.
\bibliographystyle{unsrt}
\bibliography{references}

\begin{thebibliography}{10}

\bibitem{QFT}
D.~Coppersmith.
\newblock An approximate fourier transform useful in quantum factoring.
\newblock {\em IBM Research Report}, RC 19642, 1994.

\bibitem{Kit_QPE}
A.~Yu. Kitaev.
\newblock Quantum measurements and the abelian stabilizer problem.
\newblock {\em arXiv:quant-ph/9511026}, 1995.

\bibitem{Shor_F}
Peter~W. Shor.
\newblock Polynomial-time algorithms for prime factorization and discrete
  logarithms on a quantum computer.
\newblock {\em SIAM Journal on Computing}, 26(5):1484--1509, 1997.

\bibitem{FFT}
J.W. Cooley and J.W. Tukey.
\newblock An algorithm for machine calculation of complex fourier series.
\newblock {\em Mathematics of Computation}, 19(90), 1965.

\bibitem{NielsenChua}
M.~A. Nielsen and I.~L. Chuang.
\newblock {\em Quantum Computation and Quantum Information}.
\newblock Cambridge University Press, Cambridge, UK, 2010.

\bibitem{QIP}
Glenn~J. Beach, Chris Lomont, and Charles~J. Cohen.
\newblock Quantum image processing (quip).
\newblock {\em 32nd Applied Imagery Pattern Recognition Workshop, Proceedings},
  pages 39--44, 2003.

\bibitem{QIP_2}
Fei Yan, Salvador~E. Venegas-Andraca, and Kaoru Hirota.
\newblock Toward implementing efficient image processing algorithms on quantum
  computers.
\newblock {\em Soft Computing}, 2022.

\bibitem{QIP_3}
Yue Ruan, Xiling Xue, and Yuanxia Shen.
\newblock Quantum image processing: Opportunities and challenges.
\newblock {\em Mathematical Problems in Engineering}, Vol. 2021:6671613, 2021.

\bibitem{QIP_Edge}
Xi-Wei Yao, Hengyan Wang, Zeyang Liao, Ming-Cheng Chen, Jian Pan, Jun Li,
  Kechao Zhang, Xingcheng Lin, Zhehui Wang, and Zhihuang Luo~{et al.}
\newblock Quantum image processing and its application to edge detection:
  Theory and experiment.
\newblock {\em Phys. Rev. X}, 7:031041, 2017.

\bibitem{QIP_Watermark}
Wei-Wei Zhang, Fei Gao, Bin Liu, Qiao-Yan Wen, and Hui Chen.
\newblock A watermark strategy for quantum images based on quantum fourier
  transform.
\newblock {\em Quantum Information Processing}, 12(2):793--803, 2013.

\bibitem{QCoT}
A.~Klappenecker and M.~Rotteler.
\newblock Discrete cosine transforms on quantum computers.
\newblock In {\em ISPA 2001. Proceedings of the 2nd International Symposium on
  Image and Signal Processing and Analysis. In conjunction with 23rd
  International Conference on Information Technology Interfaces (IEEE Cat.)},
  pages 464--468, 2001.

\bibitem{QIP_Inter}
Sergi Ramos-Calderer.
\newblock Efficient quantum interpolation of natural data.
\newblock {\em Phys. Rev. A}, 106:062427, 2022.

\bibitem{FFTW3}
M~Frigo and SG~Johnson.
\newblock The design and implementation of fftw3.
\newblock {\em Proceedings of the IEEE}, 93(2):216--231, 2005.

\bibitem{State_Ini}
V.V. Shende, S.S. Bullock, and I.L. Markov.
\newblock Synthesis of quantum-logic circuits.
\newblock {\em IEEE Transactions on Computer-Aided Design of Integrated
  Circuits and Systems}, 25(6):1000--1010, 2006.

\bibitem{State_Ini2}
Martin Plesch and \ifmmode \check{C}\else~\v{C}\fi{}aslav Brukner.
\newblock Quantum-state preparation with universal gate decompositions.
\newblock {\em Phys. Rev. A}, 83:032302, 2011.

\bibitem{Opt_FFT}
Lisandro Dalcin, Mikael Mortensen, and David~E. Keyes.
\newblock Fast parallel multidimensional fft using advanced mpi.
\newblock {\em Journal of Parallel and Distributed Computing}, 128:137--150,
  2019.

\bibitem{VQA_Lubasch}
Michael Lubasch, Jaewoo Joo, Pierre Moinier, Martin Kiffner, and Dieter Jaksch.
\newblock Variational quantum algorithms for nonlinear problems.
\newblock {\em Phys. Rev. A}, 101:010301, 2020.

\end{thebibliography}

\end{document}